\documentclass[]{iopart}

\usepackage{amssymb}
\usepackage{epsfig}
\usepackage{graphicx}

\begin{document}
\title{Sample dependence of the Casimir force}

\author{I. Pirozhenko$^1$, A. Lambrecht$^1$, and V. B. Svetovoy$^2$}

\address{$^1$Laboratoire Kastler Brossel, ENS, CNRS, UPMC, 4, place
Jussieu, Case 74, 75252 Paris Cedex 05, France}

\address{$^2$MESA+ Research
Institute, University of Twente, P.O. 217, 7500 AE Enschede, The
Netherlands}

\ead{V.B.Svetovoy@el.utwente.nl}


\begin{abstract}
We have analyzed available optical data for Au in the mid-infrared
range which is important for a precise prediction of the Casimir
force. Significant variation of the data demonstrates genuine sample
dependence of the dielectric function. We demonstrate that the
Casimir force is largely determined by the material properties in
the low frequency domain and argue that therefore the precise values
of the Drude parameters are crucial for an accurate evaluation of
the force. These parameters can be estimated by two different
methods, either by fitting real and imaginary parts of the
dielectric function at low frequencies, or via a Kramers-Kronig
analysis based on the imaginary part of the dielectric function in
the extended frequency range. Both methods lead to very similar
results. We show that the variation of the Casimir force calculated
with the use of different optical data can be as large as 5\% and at
any rate cannot be ignored. To have a reliable prediction of the
force with a precision of 1\%, one has to measure the optical
properties of metallic films used for the force measurement.
\end{abstract}

\maketitle

\section{Introduction\label{Sec1}}

The Casimir force \cite{Cas48} between uncharged metallic plates
attracts considerable attention as a macroscopic manifestation of
the quantum vacuum \cite{Mil94,Mos97,Mil01,Kar99,Bor01}. With the
development of microtechnologies, which routinely control the
separation between bodies smaller than 1 $\mu m$, the force became a
subject of systematic experimental investigation. Modern precision
experiments have been performed using different techniques such as
torsion pendulum \cite{Lam97}, atomic force microscope (AFM) \cite
{Moh98,Har00}, microelectromechanical systems (MEMS) \cite
{Cha01,Dec03a,Dec03b,Dec05,Ian04,Ian05} and different geometrical
configurations: sphere-plate \cite{Lam97,Har00,Dec03b}, plate-plate
\cite{Bre02} and crossed cylinders \cite{Ede00}. The relative
experimental precision of the most precise of these experiments is
estimated to be about 0.5\% for the recent MEMS measurement
\cite{Dec05} and 1\% for the AFM experiments \cite{Har00,Cha01}.

In order to come to a valuable comparison between the experiments
and the theoretical predictions, one has to calculate the force
with a precision comparable to the experimental accuracy. This is a
real challenge to the theory because the force is material, surface,
geometry and temperature dependent. Here we will only focus on the
material dependence, which is easy to treat on a level of some
percent precision but which will turn out difficult to tackle on a
high level of precision since different uncontrolled factors are
involved.

In its original form, the Casimir force per unit surface
\cite{Cas48}

\begin{equation}
F_{c}\left( a\right) =-\frac{\pi ^{2}}{240}\frac{\hbar c}{L^{4}}
\label{Fc}
\end{equation}

\noindent was calculated between ideal metals. It depends only on
the fundamental constants and the distance between the plates $L$.
The force between real materials  differs significantly
from~(\ref{Fc}) for mirror separations smaller than 1~$\mu$m.

For mirrors of arbitrary material, which can be described by
reflection coefficients, the force per unit area can be written as
\cite{Lam00}:

\begin{eqnarray}
F&=& 2\sum_{\mu}\int \frac{\mathrm{d}^{2}\mathbf{k}}{4\pi ^{2}}
\int_{0}^{\infty }\frac{\mathrm{d}\zeta }{2\pi } \hbar\kappa
\frac{r_{\mu}\left[ i\zeta ,\mathbf{k} \right]^2 e^{-2\kappa
L}}{1-r_{\mu}
\left[ i\zeta ,\mathbf{k} \right]^2 e^{-2\kappa L}}\nonumber \\
&&\kappa=\sqrt{\mathbf{k}^{2}+ \frac{\zeta ^{2}}{c^2}} \label{Force}
\end{eqnarray}

\noindent where $r_{\mu}=(r_s,r_p)$  denotes the reflection
amplitude for a given polarization $\mu=s,\;p$

\begin{eqnarray}
r_{s } &=&-\frac{\sqrt{\mathbf{k}^{2}+ \varepsilon \left( i\zeta
\right)\frac{\zeta ^{2}}{c^2}}-c\kappa } {\sqrt{\mathbf{k}^{2}+
\varepsilon \left( i\zeta \right) \frac{\zeta ^{2}}{c^2}}+c\kappa }
\nonumber \\
r_{p} &=&\frac{\sqrt{\mathbf{k}^{2}+ \varepsilon \left( i\zeta
\right)\frac{\zeta ^{2}}{c^2}}-c\kappa \varepsilon \left( i\zeta
\right) }{\sqrt{\mathbf{k}^{2}+ \varepsilon \left( i\zeta
\right)\frac{\zeta ^{2}}{c^2}}+c\kappa \varepsilon \left( i\zeta
\right) } \label{rThick}
\end{eqnarray}
The force between dielectric materials had first been derived by
Lifshitz~\cite {Lif56,LP9}. The material properties enter these
formulas via the dielectric function $\varepsilon \left( i\zeta
\right) $ at angular imaginary frequencies $\omega=i\zeta $, which
is related to the physical quantity $\varepsilon ^{\prime \prime
}\left( \omega \right)= \mathrm{Im}\left( \varepsilon \left( \omega
\right)\right) $ with the help of the dispersion relation

\begin{equation}
\varepsilon \left( i\zeta \right) -1=\frac{2}{\pi
}\int\limits_{0}^{\infty } d\omega\frac{\omega \varepsilon ^{\prime
\prime }\left( \omega \right) }{\omega ^{2}+\zeta ^{2}}.
\label{K-K}
\end{equation}

\noindent For metals $\varepsilon ^{\prime \prime }\left( \omega
\right)$ is large at low frequencies, thus the main contribution to
the integral in Eq. (\ref{K-K}) comes from the low frequencies even
if $\zeta $ corresponds to the visible frequency range. For this
reason the low-frequency behavior of $\varepsilon(\omega)$ is of
primary importance.

The Casimir force is often calculated using the optical data taken
from \cite{HB1}, which provides real and imaginary parts of the
dielectric function within some frequency range, typically between
0.1 and $10^4$~eV for the most commonly used metals, Au, Cu and Al,
corresponding to a frequency interval $[1.519\cdot 10^{14},1.519
\cdot10^{19}]$~rad/s (1~eV=$1.519 \cdot10^{15}$~rad/s \footnote{In
\protect{\cite{Lam00}} a conversion factor $1.537 \cdot
10^{15}$~rad/s was used, leading however to a negligible difference
in the Casimir force (well below 1\%).}). When the two plates are
separated by a distance $L$, one may introduce a characteristic
imaginary frequency $\zeta_{\rm ch}=c/2L$ of electromagnetic field
fluctuations in the gap.  Fluctuations of frequency $\zeta \sim
\zeta _{\rm ch}$ give the dominant contribution to the Casimir
force. For example, for a plate separation of $L=100$~nm the
characteristic imaginary frequency is $\zeta _{\rm ch}=0.988$~eV.
Comparison with the frequency interval where optical data is
available shows that the high frequency data exceeds the
characteristic frequency by 3 orders of magnitude, which is
sufficient for the calculation of the Casimir force. However, in the
low frequency domain, optical data exists only down to frequencies
which are one order of magnitude below the characteristic frequency,
which is not sufficient to evaluate the Casimir force. Therefore for
frequencies lower than the lowest tabulated frequency, $\omega _{\rm
c}$, the data has to be extrapolated. This is typically done by a
Drude dielectric function

\begin{equation}
\varepsilon \left( \omega \right) =1-\frac{\omega _{\rm
p}^{2}}{\omega \left( \omega +i\omega _{\tau }\right) },
\label{Drude}
\end{equation}

\noindent which is determined by two parameters, the plasma
frequency $\omega _{\rm p}$ and the relaxation frequency $\omega
_{\tau }$.

Different procedures to get the Drude parameters have been discussed
in the literature. They may be estimated, for example, from
information in solid state physics or extracted form the optical
data at the lowest accessible frequencies. The exact values of the
Drude parameters are very important for the precise evaluation of
the force. Lambrecht and Reynaud \cite{Lam00} fixed the plasma
frequency using the relation

\begin{equation}
\omega _{\rm p}^{2}=\frac{Ne^{2}}{\varepsilon _{0}m_{e}^{\ast }},
\label{Omp}
\end{equation}

\noindent where $N$ is the number of conduction electrons per unit
volume, $e $ is the charge and $m_{e}^{\ast }$ is the effective mass
of electron. The plasma frequency was evaluated using the bulk
density of Au, assuming that each atom gives one conduction electron
and that the effective mass coincides with the mass of the free
electron. The optical data at the lowest frequencies were then used
to estimate $\omega _{\tau }$ with the help of Eq. (\ref{Drude}). In
this way the plasma frequency $\omega _{\rm p}=9.0$~eV and the
relaxation frequency $\omega _{\tau }=0.035$~eV have been found.
This procedure was largely adopted in the following
\cite{Har00,Ede00,Cha01,Bre02,Dec03a}. However, on the example of
Cu, it was stressed in \cite{Lam00} that the optical data may vary
from one reference to another and a different choice of parameters
for the extrapolation procedure to low frequencies can influence the
Casimir force significantly.

Bostr\"om and Sernelius \cite{Bos00b} and Svetovoy and Lokhanin
\cite{Sve00b} extracted the low-frequency optical data by fitting
them with Eq. (\ref{Drude}). For one set of data from Ref.
\cite{HB2} the result \cite{Sve00b} was close to that found by the
first approach, but using different sources for the optical data
collected in Ref. \cite{HB2} an appreciable difference was found
\cite{Sve00a,Sve00b}. This difference was attributed to the defects
in the metallic films which appear as the result of the deposition
process. It was indicated that the density of the deposited films is
typically smaller and the resistivity larger than the corresponding
values for the bulk material. The dependence of optical properties
of Au films on the details of the deposition process, annealing,
voids in the films, and grain size was already discussed in the
literature \cite{Sve03b}.

In this paper we analyze the optical data for Au from several
available sources, where the mid-infrared frequency range was
investigated. The purpose is to establish the variation range of the
Drude parameters and calculate the uncertainty of the Casimir force
due to the variation of existing optical data. This uncertainty is
of great importance in view of the recent precise Casimir force
measurement \cite{Che04,Dec05} which have been performed with high
experimental accuracy. On the other hand, sophisticated theoretical
calculations predict the Casimir force at the level of 1\% or
better. These results illustrate the considerable progress achieved
in the field in only one decade. In order to assure a comparison
between theory and experiment at the same level of precision, one
has to make sure that the theoretical calculation considers
precisely the same system investigated in the experiment. This is
the key point we want to address in our paper. With our current
investigation we find an intrinsic force uncertainty of the order of
5\% coming from the fact that the Drude parameters are not precisely
known. These parameters may vary from one sample to another,
depending on many details of the preparation conditions. In order to
assure a comparison at the level of 1\% or better between
theoretical predictions and experimental results for the Casimir
force, the optical properties of the mirrors have to be measured in
the experiment.

The paper is organized as follows. In Sec. \ref{Sec2} we explain and
discuss the importance of the precise values of the Drude
parameters. In Sec. \ref{Sec3} the existing optical data for gold
are reviewed and analyzed. The Drude parameters are extracted from
the data by fitting both real and imaginary parts of the dielectric
function at low frequencies in Sec. \ref{Sec4}. In Section
\ref{Sec5} the Drude parameters are estimated by a different method
using Kramers-Kroning analysis. The uncertainty in the Casimir force
due to the sample dependence is evaluated in Sec. \ref{Sec6} and we
present our  conclusions in Sec. \ref{Sec7}.

\section{Importance of the values of the Drude parameters\label{Sec2}}
\begin{figure}[tbp]
\epsfig{file=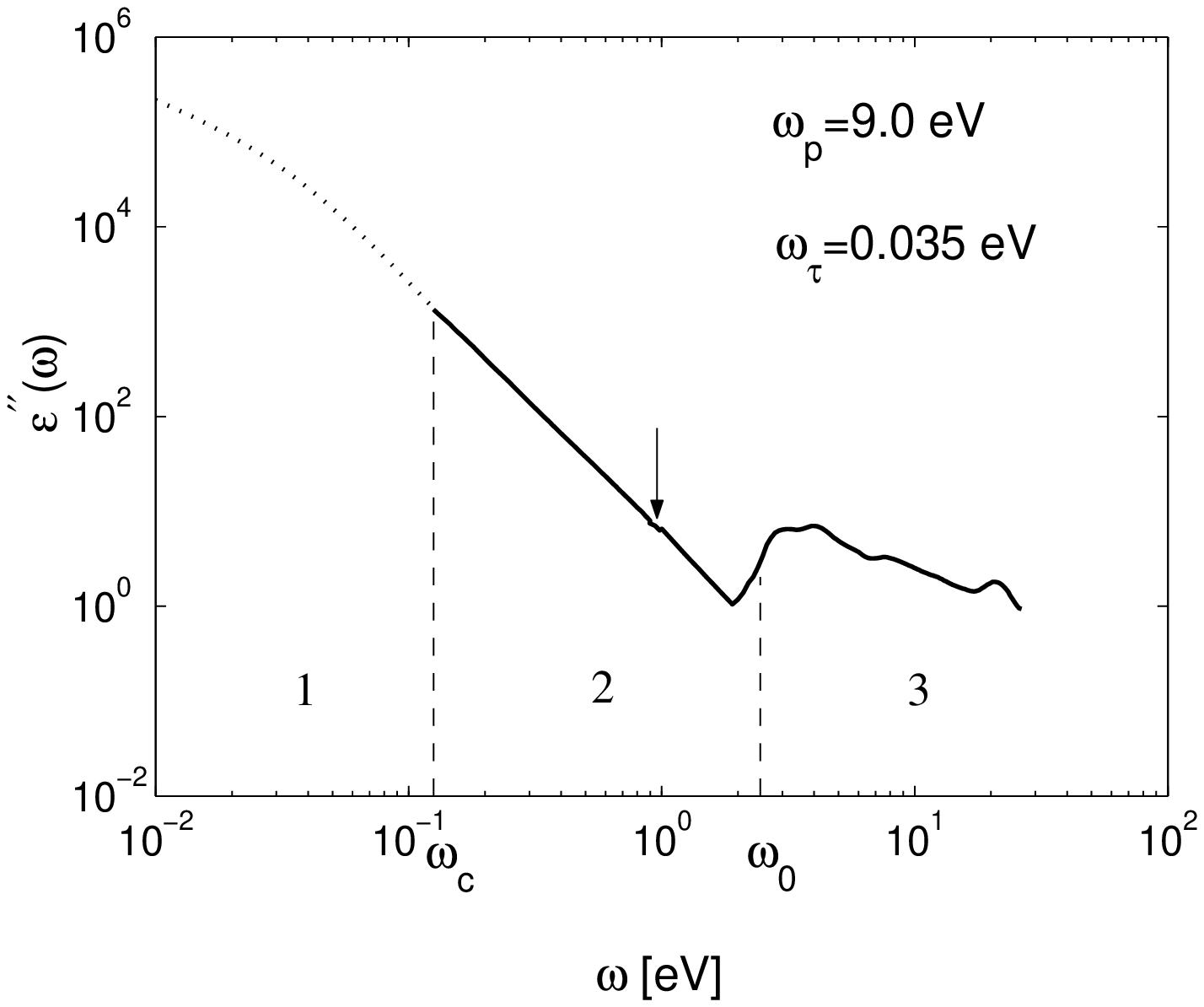,height=6cm,width=6.5cm}
\epsfig{file=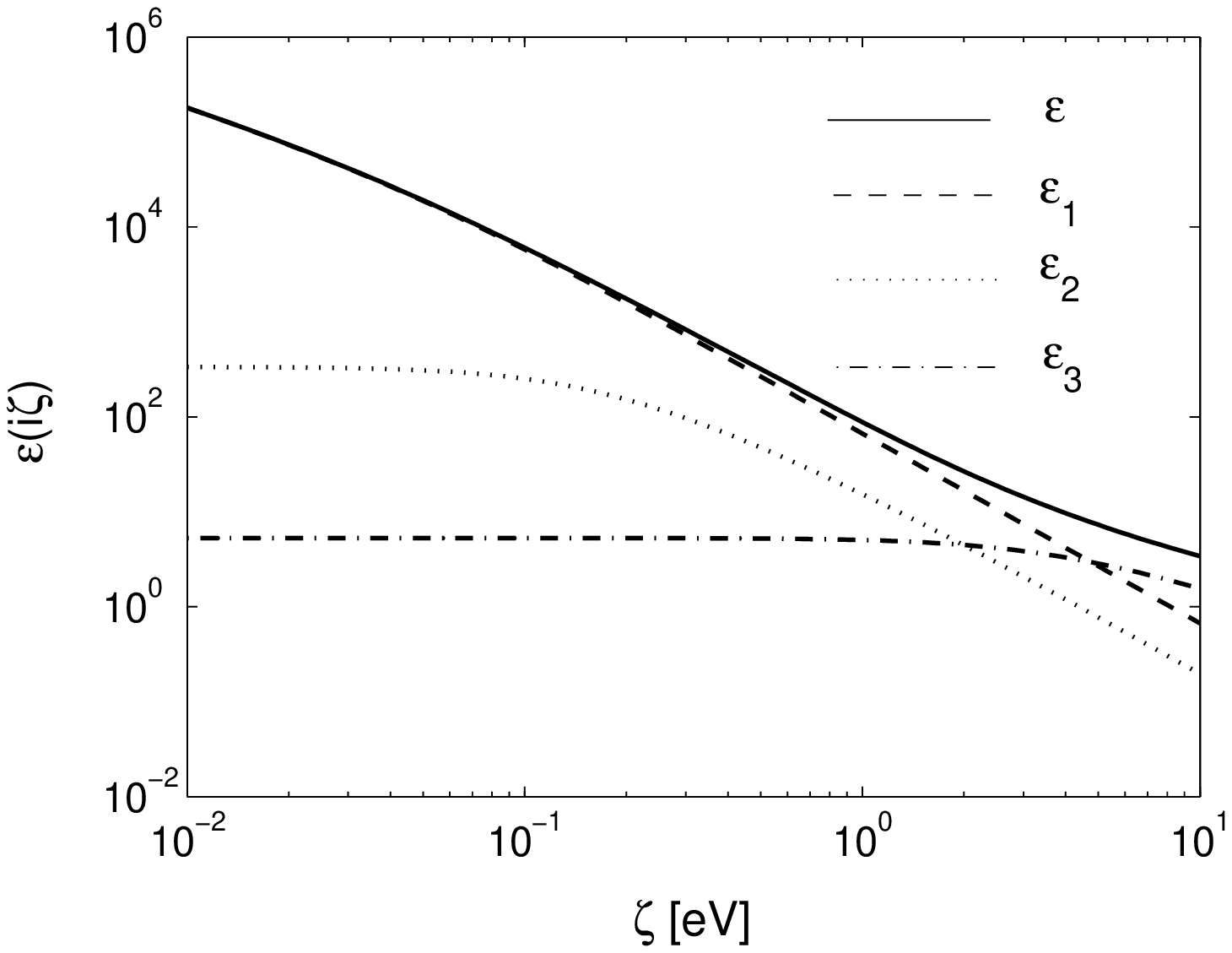,height=6cm,width=6.5cm} \caption{Left
panel: Palik's Handbook data for Au \cite{HB1} (solid line)
extrapolated to low frequencies (dotted line) with the Drude
parameters indicated in the corner. Right panel: contributions of
different real frequency domains to the dielectric function on the
imaginary axis $\varepsilon(i\zeta)$.} \label{fig1}
\end{figure}

In Figure \ref{fig1} (left) we present a typical plot of the
imaginary part of the dielectric function, which comprises Palik's
Handbook data for gold \cite{HB1}. The solid line shows the actual
data taken from two original sources: the points to the right of the
arrow are those by Th\`{e}ye \cite{The70} and to the left by Dold
and Mecke \cite{Dol65}. No data is available for frequencies smaller
than the cutoff frequency $\omega _{\rm c}$ ($0.125$~eV for this
data set) and $\varepsilon ^{\prime \prime }\left( \omega \right) $
has to be extrapolated into the region $\omega <\omega _{\rm c}$.
The dotted line shows the Drude extrapolation with the parameters
$\omega _{\rm p}=9.0$~eV and $\omega _{\tau }=0.035$~eV obtained in
Ref.~\cite{Lam00}.

One can separate three frequency regions in Fig.~\ref{fig1}~(left
panel). The region marked as {1} corresponds to the frequencies
smaller than $\omega _{\rm c}$. The region {2} defining the Drude
parameters extends from the cutoff frequency to the edge of the
interband absorption $\omega _{0}$. The high energy domain
$\omega>\omega _{0}$ is denoted by {3}.

We may now deduce the dielectric function  at imaginary
frequencies~(\ref{K-K}) using the Kramers-Kronig relation

\begin{equation}
\varepsilon \left( i\zeta \right) =1+\varepsilon _{1}\left( i\zeta \right)
+\varepsilon _{2}\left( i\zeta \right) +\varepsilon _{3}\left( i\zeta
\right) ,  \label{split}
\end{equation}

\noindent where the indices 1, 2, and 3 indicate respectively the
integration ranges $0\leq \omega <\omega _{\rm c}$, $\omega _{\rm
c}\leq\omega <\omega _{0}$, and $\omega _{0}\leq \omega <\infty $.
$\varepsilon _{1}$ can be derived using the Drude model
(\ref{Drude}) leading to

\begin{equation}
\varepsilon _{1}\left( i\zeta \right) =\frac{2}{\pi }\frac{\omega
_{p}^{2}}{\zeta ^{2}-\omega _{\tau }^{2}}\left[ \tan ^{-1}\left( \frac{\omega
_{c}}{\omega _{\tau }}\right) -\frac{\omega _{\tau }}{\zeta }\tan ^{-1}\left(
\frac{\omega _{c}}{\zeta }\right) \right] .  \label{eps1}
\end{equation}

\noindent The two other functions $\varepsilon _{2}$ and
$\varepsilon _{3}$ have to be calculated numerically. The results
for all three functions as well as for $ \varepsilon \left( i\zeta
\right) $ are shown in Fig. \ref{fig1}~(right). One can clearly see
that $\varepsilon _{1}\left( i\zeta \right) $ dominates the
dielectric function at imaginary frequencies up to $\zeta \approx
5$~eV. $\varepsilon _{2}\left( i\zeta \right) $ gives a perceptible
contribution to $\varepsilon \left( i\zeta \right)$, while
$\varepsilon_{3}\left( i\zeta \right)$ produces minor contribution
negligible for $\zeta<0.5$~eV.

As mentioned in the Introduction, we may introduce a characteristic
imaginary frequency $\zeta_{\rm ch}=c/2L$ of field fluctuations
which give the dominant contribution to the Casimir force between
two plates at a distance $L$. For a plate separation of $L=100$~nm
the characteristic imaginary frequency is $\zeta _{\rm
ch}=0.988$~eV. At this frequency the contributions of different
frequency domains to $\varepsilon \left( i\zeta _{ch}\right) $ are
$\varepsilon _{1}=68.42$, $\varepsilon _{2}=15.65$, and $\varepsilon
_{3}=5.45$. This means that for all experimentally investigated
situations, $L\gtrsim100$~nm, region {1}, corresponding to the
extrapolated optical data, gives the main contribution to
$\varepsilon \left( i\zeta \right)$. It is therefore important to
know precisely the Drude parameters.

\section{Analysis of different optical data for gold\label{Sec3}}

The optical properties of gold were extensively investigated in
50-70th. In many of those works the importance of sample preparation
methods was recognized and carefully discussed. A complete
bibliography of the publications up to 1981 can be found in Ref.
\cite{Wea81}. Regrettably the contemporary studies of gold
nanoclusters produce data inappropriate for our purposes. Among
recent experiments let us mention the measurement of normal
reflectance for evaporated gold films  \cite{Sot03}, which was
performed in the wide wavelength range $0.3-50$ $\mu$m, but
unfortunately does not permit to evaluate independently both real
and imaginary parts of the dielectric function. In contrast, the use
of new ellipsometric techniques~\cite{An02,Xia00} has produced data
for the real and imaginary part of the dielectric function for
energy intervals $1.5-4.5$~eV \cite{Wan98} and $1.5-3.5$~eV
\cite{Ben99}.

\begin{figure}[tbp]
\epsfig{file=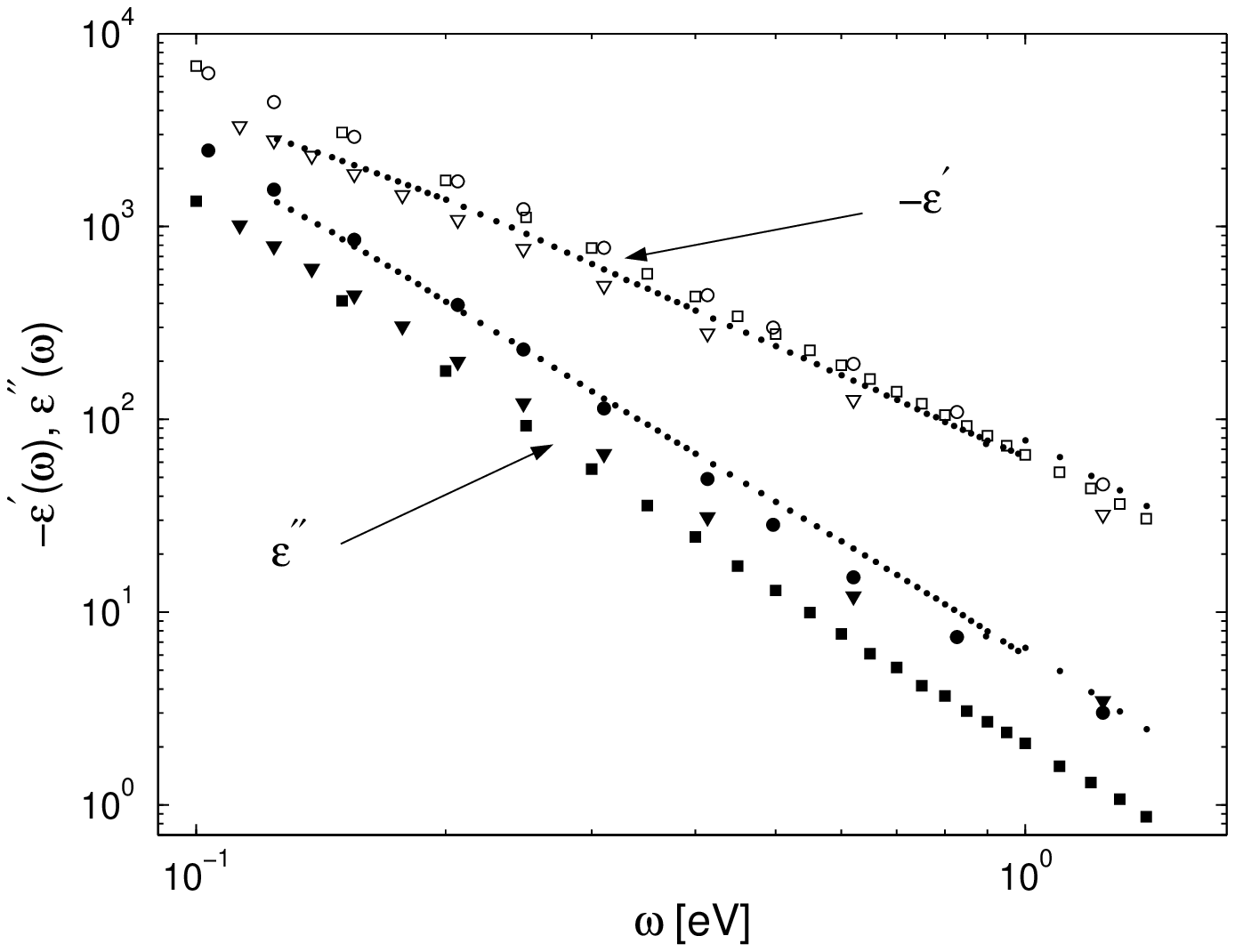,height=6cm,width=6.5cm}
\epsfig{file=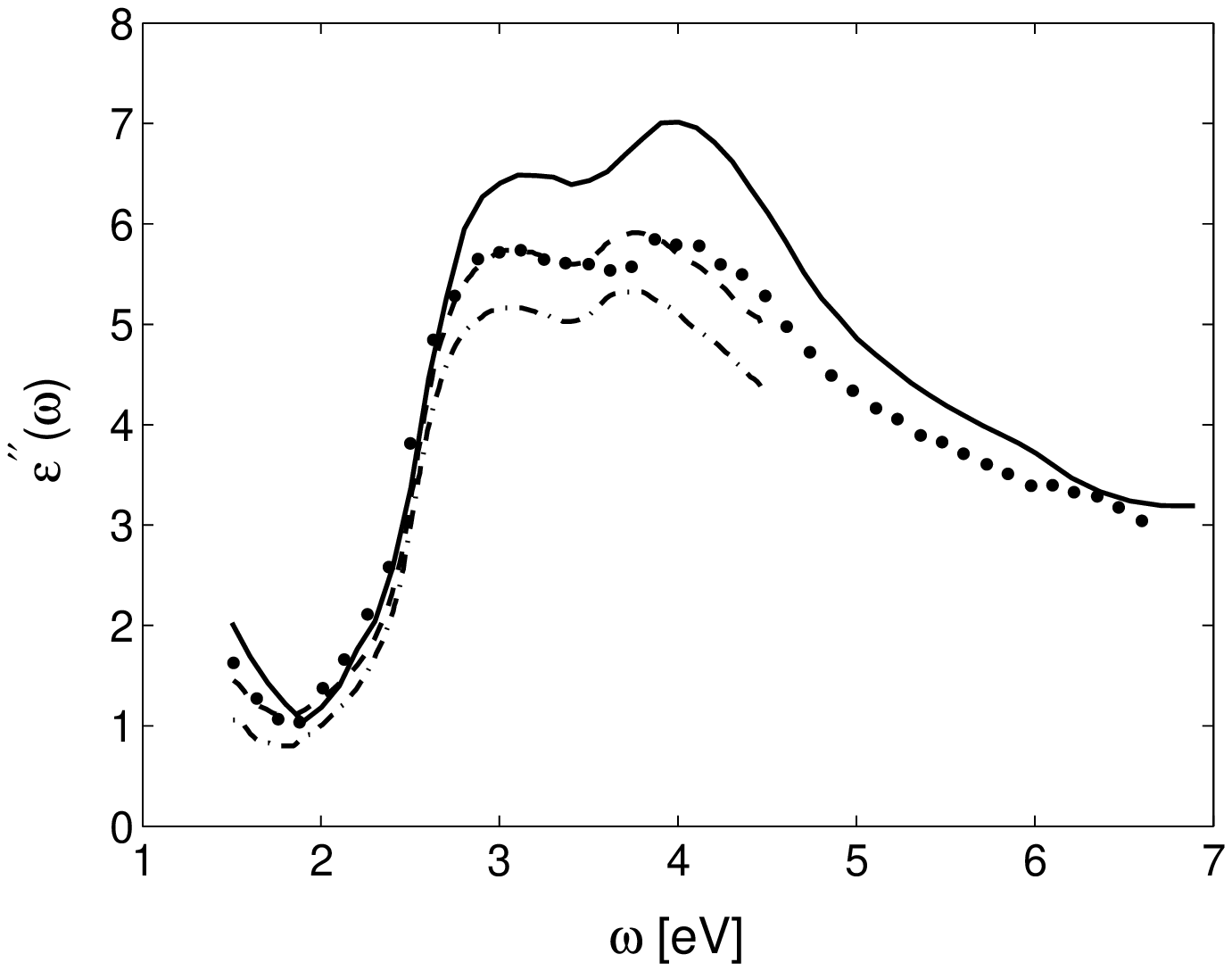,height=6cm,width=6.5cm} \caption{Left
panel: Available optical data in the mid-infrared region. The dots
represent the Dold and Mecke data for $\omega<1$~eV \cite{Dol65} and
Th\`{e}ye data \cite{The70} for higher frequencies. The squares
denote the Weaver data \cite{Wea81}. The circles stand for the data
from~\cite{Mot64}. The triangles represent the data \cite{Pad61}.
Solid squares, circles, and triangles are used to mark $ \varepsilon
^{\prime \prime }\left( \omega \right) $ while the open symbols are
used for $ \varepsilon ^{\prime }\left( \omega \right) $. Right
panel: $\varepsilon^{\prime\prime}(\omega)$ in the interband region
for different samples. The solid line represents the data measured
with the well annealed bulk-like film by Th\`{e}ye \cite{The70}. The
dots are the data by Johnson and Christy \cite{Joh72} found for
unannealed films. The dashed and dash-dotted lines are recent data
sets by Wang et al. \cite{Wan98} for unannealed films. They
correspond to films deposited with e-beam and thermal evaporation
methods, respectively. } \label{fig2}
\end{figure}

A significant amount of data in the interband absorption region
(domain {3}) has been obtained by different methods under different
conditions \cite{Pel69,The70,Joh72,Gue75,Asp80,Wan98,Ben99}. Though
this frequency band is not very important for the Casimir force, it
provides information on how the data may vary from one sample to
another.  On the contrary there are only a few sources where optical
data was collected in the mid-infrared (domain 2) and from which the
dielectric function can be extracted. The data available for $
\varepsilon ^{\prime }\left( \omega \right) $ and $ \varepsilon
^{\prime \prime }\left( \omega \right) $ in the range $\omega
<1.5$~eV and interband absorption domain {3} are presented
respectively in the left and right graph of Fig. \ref{fig2}. These
data sets demonstrate considerable variations of the dielectric
function from one sample to another.

Let us briefly discuss the  sets of
data~\cite{HB1,Wea81,Mot64,Pad61} used in our analysis and the
corresponding samples. The commonly used Handbook of Optical
Constants of Solids~\cite{HB1} comprises the optical data covering
the region from $0.125$ to $9184$~eV (dots in  Fig. \ref{fig2}). The
experimental points are assembled from several sources. For
$\omega<1$~eV they are reported by Dold and Mecke~\cite {Dol65}. For
higher frequencies up to $6$~eV they correspond to the Th\`{e}ye
data \cite{The70}. Dold and Mecke give only little information about
the sample preparation, reporting that the films were evaporated
onto a polished glass substrate and measured in air by using an
ellipsometric technique \cite{Dol65}. Annealing of the samples was
not reported.

Th\`{e}ye \cite{The70} described her films very carefully. The
samples were semitransparent Au films with a thickness of
$100-250$~\AA\ evaporated in ultrahigh vacuum on supersmooth fused
silica. The substrate was kept in most cases at room temperature.
After the deposition the films were annealed in the same vacuum at $
100-150^{\circ }$~C. The structure of the films was investigated by
X-ray and transmission-electron-microscopy methods. The dc
resistivity of the films was found to be very sensitive to the
preparation conditions. The errors in the optical characteristics of
the films were estimated on the level of a few percents.

The handbook~\cite{Wea81} embraces  the optical data from $0.1$~eV
to $28.6$~eV (marked with squares in Fig. 2). The data in the domain
$\omega<4$~eV is provided by Weaver et al. \cite{Wea81}. The values
of $\varepsilon(\omega)$ were found for the electropolished bulk
Au(110) sample. Originally the reflectance was measured in a broad
interval $0.1\leq \omega \leq 30$~eV and then the dielectric
function was determined by a Kramers-Kronig analysis. Due to
indirect determination of $\varepsilon $ the recommended accuracy of
these data sets is only 10\%.

The optical data of Motulevich and Shubin \cite{Mot64} for Au films
is marked with circles in Fig. 2. In this paper the films were
carefully described. Gold was evaporated on polished glass at a
pressure of $\sim 10^{-6}$~Torr. The investigated films were $0.5-1\
\mu$m thick. The samples were annealed in the same vacuum at
$400^{\circ }$~C for more than 3 hours. The optical constants $n$
and $k$ ($n+ik=\sqrt{\varepsilon }$) were measured by polarization
methods in the spectral range $1-12\ \mu$m. The errors in $n$ and
$k$ were estimated as 2-3\% and 0.5-1\%, respectively.

Finally, the triangles represent Padalka and Shklarevskii data \cite{Pad61}
for unannealed Au films evaporated onto glass.

The variation of the data points from different sources cannot be
explained by experimental errors. The observed deviation is the
result of different preparation procedures and reflects genuine
difference between samples. The deposition method, type of the
substrate, its temperature, quality and the deposition rate
influence the optical properties. When we are speaking about a
precise comparison between theory and experiment for the Casimir
force at the level of 1\% or better, there is no such material as
gold in general any more. There is only a gold sample prepared under
definite conditions.

\section{Evaluation of the Drude parameters through extrapolation\label{Sec4}}

We will now use the available data in the mid-infrared region to
extrapolate into the low frequency range. If the transition between
inter- and intraband absorption in gold is sharp, the data below
$\omega _{0}$ should be well described by the Drude function

\begin{equation}
\varepsilon ^{\prime }\left( \omega \right) =1-\frac{\omega
_{p}^{2}}{\omega ^{2}+\omega _{\tau }^{2} },\ \ \varepsilon ^{\prime
\prime }\left( \omega \right) =\frac{\omega _{p}^{2}\omega _{\tau
}}{\omega \left( \omega ^{2}+\omega _{\tau }^{2}\right). }
\label{ImDrude}
\end{equation}

\noindent For $\omega \gg \omega _{\tau }$, the data on the log-log
plot should fit straight lines with the slopes $-2$ and $-3$ for
$\varepsilon ^{\prime }$ and $\varepsilon ^{\prime\prime }$,
respectively, shifted along the ordinate due to variation of the
parameters for different samples. The data points in the right graph
of Fig.~\ref{fig2} are in general agreement with these expectations.
The onset values for $\varepsilon ^{\prime\prime }$,
$\ln(\omega_{\rm p}^2\omega_{\tau})$, vary more significantly due to
a significant change in $\omega_{\tau}$ for different samples, but
the Casimir force is in general not very sensitive to the relaxation
parameter \cite{Lam00}. The onset values for $-\varepsilon ^{\prime
}$, $\ln(\omega_{\rm p}^2)$, vary less but this variation is more
important for the Caimir force, which is particularly sensitive to
the value of the plasma frequency $\omega_{\rm p}$. The Drude
parameters can be found by fitting both $\varepsilon ^{\prime }$ and
$ \varepsilon ^{\prime \prime }$ with the functions (\ref{ImDrude}).
This procedure is discussed below.

The dielectric function for low frequencies, $\omega < \omega_{\rm
c}$, is found by the extrapolation of the optical data from the
mid-infrared domain, $\omega_{\rm c}<\omega<\omega_0$. The real and
imaginary parts of $\varepsilon $ follow from Eq. (\ref{ImDrude})
with an additional polarization term ${\cal P}$ in $\varepsilon
^{\prime }$:

\begin{equation}
\varepsilon ^{\prime }\left( \omega \right) ={\cal P}-\frac{\omega
_{p}^{2}}{\omega ^{2}+\omega _{\tau }^{2}},\ \ \varepsilon ^{\prime
\prime }\left( \omega \right) =\frac{\omega _{p}^{2}\omega _{\tau
}}{\omega \left( \omega ^{2}+\omega _{\tau }^{2}\right) }.
\label{DrudeRI}
\end{equation}

\noindent The polarization term appears here due to the
following reason. The total dielectric function $\varepsilon
=\varepsilon _{\left( c\right) }+\varepsilon _{\left( i\right) }$
includes contributions due to conduction electrons $\varepsilon
_{\left( c\right) }$ and the interband transitions $\varepsilon
_{\left( i\right) }$. The polarization term consists of the atomic
polarizability and polarization due to the interband transitions $
\varepsilon _{\left( i\right) }^{\prime }$

\begin{equation}
{\cal P}=1+\frac{N_{a}\alpha }{\varepsilon _{0}}+\varepsilon _{\left(
i\right) }^{\prime }\left( \omega \right) ,  \label{polariz}
\end{equation}

\noindent where $\alpha $ is the atomic polarizability and $N_{a}$
the concentration of atoms. If the transition from intra- to
interband absorption is sharp, the polarization can be considered as
constant, because the interband transitions have a threshold
behavior with an onset frequency $\omega _{0}$ and the
Kramers-Kronig relation allows one to express $\varepsilon _{\left(
i\right) }^{\prime }$ as

\begin{equation}
\varepsilon _{\left( i\right) }^{\prime }\left( \omega \right) =\frac{2}{\pi
}\int\limits_{\omega _{0}}^{\infty }dx\frac{x\varepsilon _{\left( i\right)
}^{\prime \prime }\left( x\right) }{x^{2}-\omega ^{2}}.  \label{KKi}
\end{equation}

\noindent For $\omega \ll \omega _{0}$ this integral does not depend
on $\omega $, leading to a constant $\varepsilon _{\left( i\right)
}^{\prime }\left( \omega \right) $. In reality the situation is more
complicated because the transition is not sharp and many factors can
influence the transition region. We will assume here that ${\cal P}$
is a constant but the fitting procedure will be shifted to
frequencies where the transition tail is not very important. In
practice Eq. (\ref{DrudeRI}) can be applied for $\omega <1$ eV.

Our purpose is now to establish the magnitude of the force change
due to reasonable variation of the optical properties. To this end
the available low-frequency data for $\varepsilon ^{\prime }\left(
\omega \right) $ and $\varepsilon ^{\prime\prime }\left( \omega
\right) $ presented in the left graph of Fig. \ref{fig2} were fitted
with Eq. (\ref{DrudeRI}). The results together with the expected
errors are collected in Table \ref{tab1}.

\begin{table}
\centering
\begin{tabular}{l||l|l|l|l}
N & $\ \ \ \omega _{p}$(eV) & $\omega _{\tau }\cdot 10^{2}$(eV) & \
\ \ \ \ ${\cal P}$ \\ \hline\hline 1 & $7.50\pm 0.02$ &
$6.1\pm 0.07$ & $-27.67\pm 5.79$ & Palik, 66 points , $\ \cdot$ \\
2 & $8.41\pm 0.002$ & $2.0\pm 0.005$ & $7.15\pm 0.035$  &
Weaver, 20 points, \ $\blacksquare, \Box $\\
3 & $8.84\pm 0.03$ & $4.2\pm 0.06$ & $12.94\pm 16.81$  &
Motulevich, 11 points, \ $\bullet, \circ$\\
4 & $6.85\pm 0.02$ & $3.6\pm 0.05$ & $-12.33\pm 9.13$ & Padalka 11
points, \ $\blacktriangledown,\triangledown$
\end{tabular}
\caption{The Drude parameters found by fitting the available
infrared data for $\varepsilon ^{\prime }\left( \omega \right)$ and
$\varepsilon ^{\prime \prime }\left( \omega \right) $ with Eq.
(\ref{DrudeRI}). The error is statistical.}\label{tab1}
\end{table}

The error in Table \ref{tab1} is the statistical uncertainty. It was
found using a $\chi ^{2}$ criterion for joint estimation of 3
parameters \cite{PatDat}. For a given parameter the error
corresponds to the change $\Delta\chi ^{2}=1$ when two other
parameters  are kept constant. The parameter ${\cal P}$
enters~(\ref{DrudeRI}) as an additive constant and in the considered
frequency range its value is smaller than  1\% of $\varepsilon
^{\prime }\left( \omega \right)$ . That is why the present fitting
procedure cannot  resolve it with reasonable errors.

As mentioned before, in the case of the Weaver data \cite{Wea81} the
recommended precision in $\varepsilon^{\prime}$ and
$\varepsilon^{\prime\prime}$ is 10\% while Motulevich and Schubin
reported 2-3\% and 0.5-1\% errors in $n$ and $k$. We did not take
these errors explicitly into account as we do not know if they are
of statistical or systematic nature or a combination of both. But to
illustrate their possible influence let us just mention that if we
interpret them as systematic errors, we can propagate the errors in
$\varepsilon$ or $n,k$ to the values of $\omega_{\rm p}$ and
$\omega_{\tau}$, leading to an additional error in $\omega_{\rm p}$
of about 5\% for the Weaver data and 1\% for the Motulevich data and
twice as large in $\omega_{\tau}$.

Significant variation of the plasma frequency, well above the
errors, is a distinctive feature of the table. The bulk and annealed
samples (rows 2 and 3) demonstrate larger values of $\omega _{\rm
p}$. The rows 1 and 4 corresponding to the evaporated unannealed
films give rise to considerably smaller plasma frequencies $\omega
_{\rm p}$. Note that our calculations are in agreement with the one
given by the authors \cite{Dol65,Pad61} themselves.

\begin{figure}[tbp]
\epsfig{file=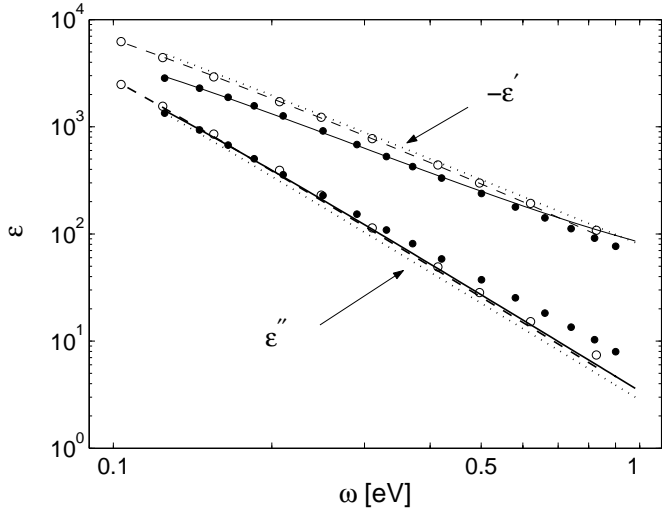,width=9cm}\newline \caption{The infrared
optical data by Dold and Mecke \cite{Dol65} (full circles) and by
Motulevich and Shubin \cite{Mot64} (open circles) together with the
best Drude fits given by the solid and dashed lines, respectively.
The dotted lines present the fit with $\omega_{\rm p}=9$ eV and
$\omega_{\tau}=35$ meV which agrees better with the Motulevich and
Shubin data (open circles) than with the handbook data (full
circles). } \label{fig5}
\end{figure}

To have an idea of the quality of the fitting procedure, we show in Fig.
\ref{fig5} the experimental points and the best fitting curves
for Dold and Mecke data \cite{Dol65,HB1} (full circles and solid
lines) and Motulevich and Shubin data \cite{Mot64} (open circles and
dashed lines). Only 25\% of the points from \cite{HB1} are shown for
clarity. One can see that for $\varepsilon ^{\prime \prime }$ at
high frequencies the dots lie above the solid line demonstrating
presence of a wide transition between inter- and intraband absorption.
Coincidence of the solid and dashed lines for $\varepsilon ^{\prime
\prime }$ is accidental. The fits for $\varepsilon ^{\prime }$ are
nearly perfect for both data sets.

It is interesting to see on the same figure how well the parameters
$\omega _{\rm p}=9.0$ eV, $\omega _{\tau }=0.035$ eV agree with the
data in the mid-infrared range.  The curves corresponding to this
set of parameters are shown in Fig. \ref{fig5} as dotted lines. One
can see that the dotted line, which describes $\varepsilon ^{\prime
\prime }$ is very close to the solid line. However, the dotted line
for $ \varepsilon ^{\prime }$ does not describe well the handbook
data (full circles). It agrees much better with Motulevich and
Shubin data \cite{Mot64} (open circles). The reason for this is that
$\omega _{\rm p}=9.0$ eV is the maximal plasma frequency for Au. Any
real film may contain voids leading to smaller density of electrons
and, therefore, to smaller $\omega _{\rm p}$. Motulevich and Shubin
\cite{Mot64} annealed their films which reduced the number of
defects and made the plasma frequency close to its maximum. A plasma
frequency $\omega _{\rm p}=9.0$ eV was also reported in Ref.
\cite{Ben66}, where the authors checked the validity of the Drude
theory by measuring reflectivity of carefully prepared gold films in
ultrahigh vacuum in the spectral range $0.04<\omega<0.6$ eV.
Therefore, this value is good if one disposes of well prepared
samples.

\section{The Drude parameters from Kramers-Kronig analysis\label{Sec5}}

Because the values of the Drude parameters are crucial for a
reliable prediction of the Casimir force, it is important to assess
that different methods to determine the parameters give the same
results. Alternatively to the extrapolation procedure of the
previous section  we will now discuss a procedure based on a
Kramers-Kronig analysis. To this aim we will extrapolate only the
imaginary part of the dielectric function to low frequencies
$\omega<\omega_{\rm c}$. The dispersion relation between
$\varepsilon^{\prime}$ and $\varepsilon^{\prime\prime}$

\begin{equation}\label{KKrel}
    \varepsilon^{\prime}(\omega)-1=\frac{2}{\pi
}P\int\limits_{0}^{\infty }dx\frac{x\varepsilon ^{\prime \prime
}\left( x\right) }{x^{2}-\omega ^{2}}
\end{equation}

\noindent can then be used to predict the behavior of
$\varepsilon^{\prime}(\omega)$ and compare it with the one observed in
the experiments. From this comparison the Drude parameters can be
extracted.

The low-frequency behavior of $\varepsilon^{\prime\prime}(\omega)$
is important for the prediction of $\varepsilon^{\prime}$ because
for metals $\varepsilon^{\prime\prime}(\omega)\gg1$ in the low
frequency range. Therefore, at $\omega<\omega_{\rm c}$ we are using
$\varepsilon^{\prime\prime}(\omega)$ from Eq. (\ref{ImDrude}). At
higher frequencies the experimental data from different sources
\cite{HB1,Wea81,Mot64,Pad61} are used. The data in Refs.
\cite{Mot64,Pad61} must be extended to high frequencies starting
from $\omega=1.25$~eV. We do this using the handbook data
\cite{HB1}.

Let us start from the data for bulk Au(110) \cite{Wea81}. This data
set is given in the interval $0.1<\omega<30$~eV. Below
$\omega=0.1$~eV we use the Drude model for
$\varepsilon^{\prime\prime}$ and above $\omega=30$~eV the cubic
extrapolation $C/\omega^3$. The Drude parameters are practically
insensitive to the high frequency extrapolation. The data set was
divided into overlapping segments containing 12 points. Each segment
was fitted with a polynomial of forth order in frequency. The first
segment, were $\varepsilon^{\prime\prime}(\omega)$ increases very
fast, was fitted with the polynomial in $1/\omega$. Then, in the
range of overlap (4 points) a new polynomial smoothly connecting two
segments was chosen. In this way we have fitted the experimental
data with a function which is smooth up to the first derivative.

\begin{figure}[tbp]
\epsfig{file=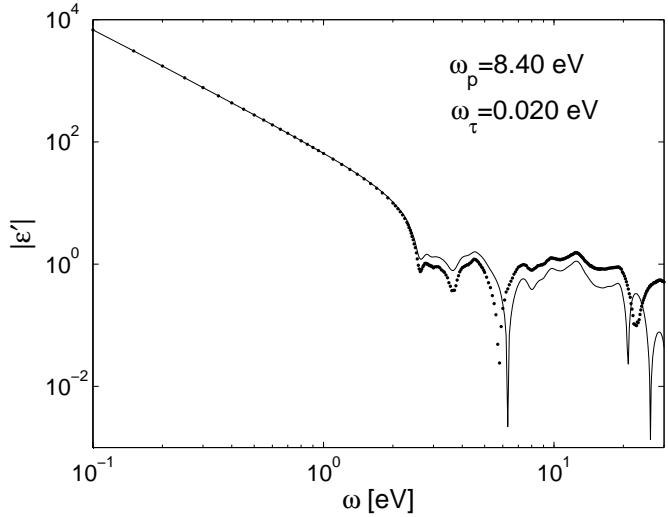,width=9cm}\newline
\caption{$|\varepsilon^{\prime}|$ as a function of $\omega$ for bulk
gold. Dots are the experimental data \cite{Wea81}. The solid line is
the prediction according to Eq. (\ref{KKrel}) with the Drude
parameters $\omega_p=8.40\ eV$, $\omega_{\tau}=0.02\ eV$. }
\label{fig6}
\end{figure}

The real part of the dielectric function
$\varepsilon^{\prime}(\omega)$ is predicted by Eq. (\ref{KKrel}) as
a function of the Drude parameters $\omega_p$ and $\omega_{\tau}$.
These parameters are chosen such as to minimize the difference
between observed and predicted values of
$\varepsilon^{\prime}(\omega)$, leading to $\omega_{\rm p}=8.40$~eV
and $\omega_{\tau}=0.020$~eV. These parameters are in reasonable
agreement with the ones indicated in Tab. \ref{tab1}. In Fig.
\ref{fig6} the experimental data (dots) and
$|\varepsilon^{\prime}(\omega)|$ found from Eq. (\ref{KKrel}) (solid
line) are plotted, showing perfect agreement at low frequencies,
while at high frequencies $\omega>2.6$~eV the agreement is not very
good. This may be fixed by choosing an appropriate high frequency
extrapolation. We do not give these details here as this
extrapolation has practically no influence on the Drude parameters.

\begin{figure}[tbp]
\epsfig{file=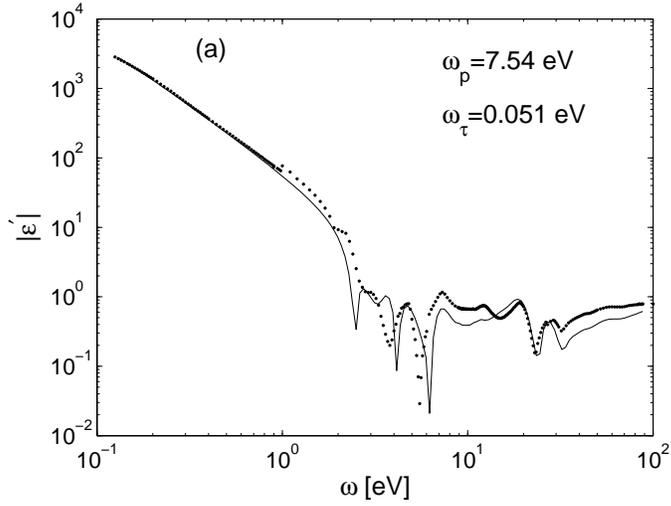,width=9cm}\newline
\caption{$|\varepsilon^{\prime}|$ as a function of $\omega$ for
handbook data \cite{HB1} (dots). The solid line is found from
Kramers-Kronig relation. The Drude parameters correspond to minimal
deviations between experimental data and calculations.} \label{fig7}
\end{figure}

When applying the same procedure to the handbook data \cite{HB1}, we
find $\omega_p=7.54$~eV and $\omega_{\tau}=0.051$~eV, again in
agreement with the parameters indicated in Tab.~\ref{tab1}.
Fig.~\ref{fig7} shows a plot of $\varepsilon^{\prime}(\omega)$
predicted with these parameters. At low frequencies the agreement
with the experimental data is good but it becomes worse when the
interband data \cite{Dol65} joins the intraband (high frequency)
data \cite{The70}. These two data sets correspond to samples with
different optical properties. In this case the dispersion relation
(\ref{KKrel}) is not necessarily very well satified. In contrast
with the previous case, high frequency extrapolation cannot improve
the situation; it influences the curve only marginally.

\begin{figure}[tbp]
\epsfig{file=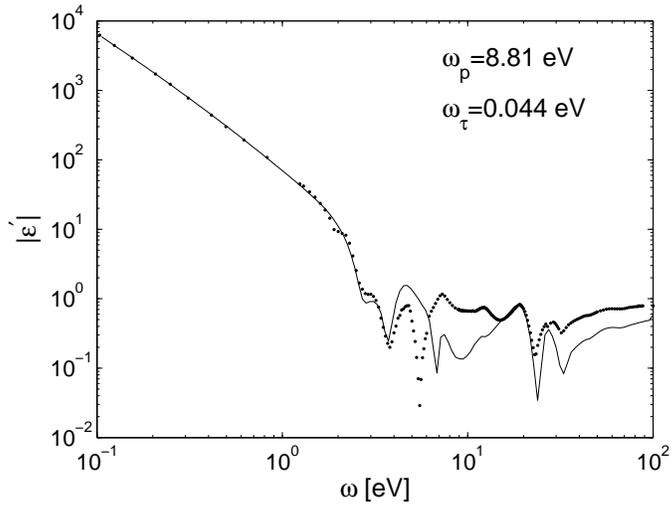,width=9cm}
\newline
\caption{$|\varepsilon^{\prime}|$ as a function of $\omega$ for
Motulevich and Shubin data \cite{Mot64} extended by the handbook
data \cite{HB1} for $\omega>1.25$~eV (dots). The solid line is found
from Kramers-Kronig relation. } \label{fig8}
\end{figure}

Following the same procedure for the Motulevich and Shubin data
\cite{Mot64}, we find the Drude parameters $\omega_{\rm p}=8.81$~eV,
$\omega_{\tau}=0.044$~eV which are close to the values in
Tab.~\ref{tab1}. The experimental data and calculated function
$|\varepsilon^{\prime}(\omega)|$ are shown in Fig.~\ref{fig8}. There
is good agreement for frequencies $\omega<4$~eV as the data in Ref.
\cite{Mot64} matches very well the Th\`{e}ye data \cite{The70}.
Deviations at higher frequencies are again quite sensitive to
high-frequency extrapolation as already noted before.

Similar calculations done for the Padalka and Shklyarevskii data
\cite{Pad61} give the Drude parameters $\omega_{\rm p}=6.88$~eV and
$\omega_{\tau}=0.033$~eV, producing good agreement only in the range
$\omega<1.3$~eV because this data set matches only poorly the
Th\`{e}ye data \cite{The70}.

Using the Kramers-Kronig analysis for the determination of the Drude
parameters leads essentially to the same parameters for all 4 sets
of the experimental data. Experimental and calculated curves for
$\varepsilon^{\prime}(\omega)$ are in very good agreement at low
frequencies. At high frequencies the agreement is not so good for
two different reasons. First, at high frequencies the calculated
curve is sensitive to the high-frequency extrapolation and thus a
better choice of this extrapolation can significantly reduce high
frequency deviations. The other reason is that one has to combine
the data from different sources to make a Kramers-Kronig analysis
possible. These data sets do not always match each other well as it
is for example the case of the Dold and Mecke data and the Th\`{e}ye
data. In this case significant errors might be introduced in the
dispersion relation. Indeed the Kramers-Kronig analysis is a
valuable tool only for data taken from the same sample.

\section{Uncertainty in the Casimir force due to variation of optical properties\label{Sec6}}

We will now assess how the values of the Casimir force are
influenced by the different values of the Drude parameters. As an
example we consider as input the optical data for Au from
\cite{HB1}.

Instead of calculating the absolute value of the Casimir force, we
will give the factor which measures the reduction of the Casimir
force with respect to the ideal Casimir force between perfect
mirrors as introduced in \cite{Lam00}

\begin{equation}\label{eta}
\eta_F=\frac{120 L^4}{c\pi^4}\int\limits_0^{\infty}d\kappa\,\kappa^2
\int\limits_0^\kappa
d\zeta\sum_{\mu}\frac{r_{\mu}^2}{e^{2\kappa}-r_{\mu}^2},
\end{equation}

\noindent The dielectric function at imaginary frequencies
$\varepsilon(i\zeta)$ is calculated using the Kramers-Kronig
relation~(\ref{K-K}) and the integration region is divided in two
parts

\begin{equation}\label{imfreq}
\int_0^{\infty}\frac{x\,
\varepsilon''(x)}{x^2+\omega^2}dx\rightarrow
\left\{\int_{0}^{x_c}+\int_{x_c}^{x_{max}}\right\}\frac{x\,
\varepsilon''(x)}{x^2+\omega^2}dx=I_1+I_2.
\end{equation}

\noindent We assume that for  $x<x_{\rm c}$ the Drude
model~(\ref{ImDrude}) is applicable. Then the integration in $I_1$
may be carried out explicitly, see~(\ref{eps1}). In $I_2$ we
integrate from $x_{\rm c}=0.125$~eV to $x_{\rm max}=9000$~eV
(corresponding to the range of available optical data in
\cite{HB1}).

For the calculation of the reduction factor (\ref{eta}) the
integration range was chosen as $10^{-4}-10^{3}$~eV. We also  varied
the integration range by half an order of magnitude, which changed
the result by less than $0.1\%$. The results of the numerical
integration are collected in Table~\ref{Tab3}.

\begin{table}
\begin{tabular}{l|l|c|c|c|c|c}
\hline &$\omega_p, \omega_{\tau}(eV)\,\backslash L (\mu m)$&$ \quad
0.1 \quad $& $\quad 0.3\quad $ &$\quad 0.5 \quad$& $\quad 1.0 \quad
$& $\quad 3.0  \quad $
\\
\hline
\hline
1.&$\omega_{\rm p}=7.50$, $\omega_{\tau}=0.061$& 0.43& 0.66 & 0.75 & 0.85  & 0.93 \\[3mm]
\hline
2.&$\omega_{\rm p}=8.41$, $\omega_{\tau}=0.02$& 0.45&  0.69 &0.79 &0.88   &0.95  \\[3mm]
\hline
3.&$\omega_{\rm p}=8.84$, $\omega_{\tau}=0.0422$& 0.46 &  0.69 & 0.78  & 0.87 &0.94 \\[3mm]
\hline
4.&$\omega_{\rm p}=6.85$, $\omega_{\tau}=0.0357$& 0.42 & 0.65  & 0.75  & 0.84 &0.93 \\[3mm]
\hline
5.&$\omega_{\rm p}=9.00$, $\omega_{\tau}=0.035$& 0.47&  0.71 & 0.79 &  0.88 & 0.95 \\[3mm]
\hline
6.&$\omega_{\rm p}=7.50\pm15\%$& 0.45&  0.68 &  0.77 &  0.86& 0.94\\
&$\omega_{\tau}=0.061$&0.41 & 0.63 & 0.73 & 0.83 & 0.92 \\[3mm]
\hline
7.&$\omega_{\rm p}=7.50$& 0.42&  0.65 &  0.74 &  0.84& 0.92\\
&$\omega_{\tau}=0.061\pm30\%$& 0.44& 0.67 & 0.76 & 0.86  & 0.93 \\
\hline
\end{tabular}
\caption{The reduction factors at different plate separations
calculated with the different pairs of values of the Drude
parameters corresponding to different data. The last two rows show
the variation of the reduction factor when either the plasma
frequency or the relaxation parameter is varied.}\label{Tab3}
\end{table}

The first four rows of the table present the reduction factors for
four pairs of the Drude parameters that were obtained by fitting the
optical data from different sources. The next row shows the result
obtained for $\omega_{\rm p}=9$~eV and $\omega_{\tau}=35$~meV. The
last two rows show the variation of the reduction factor if the
plasma frequency $\omega_{\rm p}$ or the relaxation parameter
$\omega_{\tau}$ are varied by $\pm 15\%$ and $\pm 30\%$,
respectively. The upper (lower) line corresponds here to the upper
(lower) sign.

The variation of the optical data and the associated Drude
parameters introduces a variation in the Casimir force ranging from
5.5\% at short distances (100~nm) to 1.5\% at long distances
(3~$\mu$m). The distance dependence is of course related to the fact
that the material properties influence the Casimir force much more
at short than at long plate separation. The strongest variation of
5.5\% gives an indication of the genuine sample dependence of the
Casimir force. For this reason it is necessary to measure the
optical properties of the plates used in the Casimir force
measurement if a precision of the order 1\% or better in the
\textit{comparison} between experimental values and theoretical
predictions is aimed at. Incidentally let us notice that the plasma
frequency $\omega_{\rm p}=7.5$~eV, which is found here to fit best
Palik's handbook data \cite{HB1}, is basically the same as  the one
proposed alternatively in \cite{Lam00} for Cu, which has very
similar optical properties to Au concerning the Casimir force
\cite{PRLComment}. For Cu, the variation of the plasma frequency
from $\omega_{\rm p}=9$~eV to $\omega_p=7.5$~eV introduced a
variation of the Casimir force up to 5\% \cite{Lam00}.

In order to asses more quantitatively the role of the two Drude
parameters, we show in the last two rows of table \ref{Tab3} the
variation of the reduction factor when either the plasma frequency
or the relaxation parameter is varied with the other parameter kept
constant. One can see that the increase (decrease) of the relaxation
parameter by $\delta\omega_{\tau}=30\%$ lowers (increases) the
reduction factor $\eta_F$ at $L=0.1~\mu m$  by only
$\delta\eta_F=1.6\%$.  However, the $15\%$ variation of the  plasma
frequency leads to $4.2\%$ change in the  reduction factor. Thus the
Casimir force is much more sensitive to the variation of the plasma
frequency, basically as the plasma frequency determines the
reflection quality of the plates (an infinite plasma frequency
corresponds to perfectly reflecting mirrors).

\section{Conclusions\label{Sec7}}

In this paper we have performed the first systematic and detailed
analysis of optical data for Casmir force measurements. We have
studied the relative importance of the different frequency regions
for the Casimir force as a function of the plate separation and
established the critical role of the Drude parameters in particular
for short distance measurements. We have then analyzed and compared
four different sets of optical data. For each set we have extracted
the corresponding plasma frequency and relaxation parameter either
by fitting real and imaginary part of the dielectric function at low
frequencies or by using a detailed Kramers-Kronig analysis. Both
methods lead essentially to the same results. The Kramers-Kronig
analysis reveals itself to be a powerful tool for the estimation of
the low frequency Drude parameters for data coming from the same
sample.

A variation of the values of the Casimir force up to 5.5\% is found
for different optical data sets. This gives an intrinsic unknown
parameter for the Casimir force calculations and demonstrates the
genuine sample dependence of the Casimir force. The today existing
numerical and analytical calculations of the Casimir force in
themselves are very precise. In the same way, measurements of the
Casimir force have achieved high accuracy over the last decade. In
order to compare the results of the achievements in theory and
experiment at a level of 1\% precision or better, the crucial point
is to make sure that calculations and experiments are performed for
the same physical sample. One therefore has to know the optical and
material properties of the sample used in the experiment. These
properties must be measured for frequencies as low as possible. In
practice, the material properties have to be known over an interval
of about  4 orders of magnitude around the characteristic frequency
$\zeta_{\rm ch}=c/2L$. For a plate separation of $L=100$~nm this
means an interval [10~meV, 100~eV]. If measurements at low
frequencies are not possible, the low frequency Drude parameters
should be extracted from the measured data, by one of the two
methods discussed here.

\textbf{Acknowledgements} Part of this work was funded by the
European Contract STRP 12142 NANOCASE. We wish to thank S. Reynaud
and A. Krasnoperov for useful discussions.

\end{document}